# Dirac equation with coupling to 1/r singular vector potentials for all angular momenta

A. D. Alhaidari[†]

*Saudi Center for Theoretical Physics, Dhahran, Saudi Arabia*
AND
*Physics Department, King Fahd University of Petroleum & Minerals, Dhahran 31261, Saudi Arabia*

We consider the Dirac equation in 3+1 dimensions with spherical symmetry and coupling to 1/r singular vector potential. An approximate analytic solution for all angular momenta is obtained. The approximation is made for the 1/r orbital term in the Dirac equation itself not for the traditional and more singular $1/r^2$ term in the resulting second order differential equation. Consequently, the validity of the solution is for a wider energy spectrum. As examples, we consider the Hulthén and Eckart potentials.

PACS numbers: 03.65.Pm, 03.65.Ge, 03.65.Ca

Keywords: Dirac equation, singular potentials, non-zero angular momentum, energy spectrum, Hulthén, Eckart.

Most of the three dimensional problems in quantum mechanics, other than the Coulomb and oscillator, are exactly solvable for zero angular momentum (S-wave solution). Examples include, but not limited to, potentials like the Morse, Pöschl-Teller, Hulthén, Scarf, Rosen-Morse, etc. [1]. Therefore, one looks for a numerical solution. With the considerable advancements in computing facilities, these solutions are usually very satisfactory. However, still remains missing is the power of analytic assessment in giving a deeper insight into the solution of the problem. This is especially true in certain limits where the potential parameters assume values near critical transitions. Very often in such cases, numerical computations experience some degree of instability. Therefore, many attempts have been made to give an analytic approximation of the solutions of these problems for non-zero angular momentum. A widely known example, is the Rekeris approximation [2] in which the orbital term $\ell(\ell+1)r^{-2}$ is expanded in terms of singular functions of $e^{-r/a}$ compatible with the solvability of the problem for $r \ll a$. Now, since the orbital term $r^{-2}$ is too singular, the validity of such approximations is limited only to very few of the lowest energy states. Therefore, to go to higher energy states one may attempt to solve the relativistic version of these problems, like the Dirac equation or the Klein-Gordon equation. Recently, there is a noticeable surge of interest in this problem [3]. It is unfortunate that in all such attempts full advantage of the unique features of the Dirac equation is not utilized. For example, the analytic approximation is still devoted to the new orbital term $\kappa(\kappa \pm 1)r^{-2}$ in the second order differential equation that results from the Dirac equation, where $\kappa$ is the spin-orbit quantum number. Moreover, in almost all of this type of work, "spin" or pseudo-spin" symmetry [3,4] is imposed where the Dirac spinor or the Klein-Gordon scalar particle is coupled not only to the vector potential but also to a scalar potential such that their sum or difference is a constant. Moreover, in few of these studies, the scalar potential is introduced by way of making the mass spatially

---
[†] Present temporary address: 1300 Midvale Ave. #307, Los Angeles, CA 90024

–1–

dependent [5] since the mass term in the Dirac and Klein-Gordon equations is a scalar object.

In this work, however, we perform the analytic approximation of the orbital term in the Dirac equation itself, which is a first order differential equation, not in the resulting second order differential equation. The advantage is that in such case the orbital term is less singular since it goes like $r^{-1}$ not like $r^{-2}$. However, we deploy an approach that limits our investigation to 1/r singular potentials. This, of course, is not the only method in the treatment the problem. As examples, we apply our approach to the Hulthén [6] and Eckart potentials [7]. To the best of our knowledge, this work represents the first attempt at a solution of the Dirac equation with coupling to 1/r singular *pure vector* potentials for all angular momenta, other than the Dirac-Coulomb.

In the relativistic units $\hbar = c = 1$, the Dirac equation with spherically symmetric coupling to scalar and vector potentials, $U(r)$ and $V(r)$, has the following radial component [8]

$$\begin{pmatrix} m+U(r)+V(r)-\varepsilon & -\dfrac{d}{dr}+\dfrac{\kappa}{r} \\ \dfrac{d}{dr}+\dfrac{\kappa}{r} & -m-U(r)+V(r)-\varepsilon \end{pmatrix} \begin{pmatrix} \psi^+(r) \\ \psi^-(r) \end{pmatrix} = 0. \qquad (1)$$

We approximate the $r^{-1}$ orbital term by a singular function $W(r)$ under certain approximation condition that will be *maintained* throughout the work. In that case, Eq. (1) becomes

$$\begin{pmatrix} m+U+V-\varepsilon & -\dfrac{d}{dr}+\kappa W \\ \dfrac{d}{dr}+\kappa W & -m-U+V-\varepsilon \end{pmatrix} \begin{pmatrix} \psi^+ \\ \psi^- \end{pmatrix} = 0. \qquad (2)$$

Applying the global unitary transformation $\exp\left(\frac{i}{2}\theta\sigma_2\right)$, where $\sigma_2 = \begin{pmatrix} 0 & -i \\ i & 0 \end{pmatrix}$ and $\theta$ is a constant angular parameter, takes this matrix equation into

$$\left[(m+U)\begin{pmatrix} C & -S \\ -S & -C \end{pmatrix} + \kappa W \begin{pmatrix} S & C \\ C & -S \end{pmatrix} + \begin{pmatrix} V-\varepsilon & -\dfrac{d}{dr} \\ \dfrac{d}{dr} & V-\varepsilon \end{pmatrix}\right] \begin{pmatrix} \phi^+ \\ \phi^- \end{pmatrix} = 0, \qquad (3)$$

where $S = \sin\theta$, $C = \cos\theta$ and $\begin{pmatrix} \phi^+ \\ \phi^- \end{pmatrix} = \begin{pmatrix} \cos\frac{\theta}{2} & \sin\frac{\theta}{2} \\ -\sin\frac{\theta}{2} & \cos\frac{\theta}{2} \end{pmatrix}\begin{pmatrix} \psi^+ \\ \psi^- \end{pmatrix}$. The choice $CU(r) + \kappa SW(r) = \pm V(r)$ makes one of the diagonal elements in the matrix wave operator constant. This results in a Schrödinger-like equation for one of the two spinor components that may bring about an analytic solution. In this work we specialize to pure vector potential coupling (i.e., $U = 0$) and look for problems with $V(r) = \mu W(r)$, where $\mu$ is a dimensionless physical parameter. Therefore, we can only treat problems with 1/r singular vector potentials since $W \approx 1/r$. Choosing the angular transformation parameter $\theta$ such that $\sin\theta = \pm\mu/\kappa$, where $-\frac{\pi}{2} \leq \theta \leq +\frac{\pi}{2}$, gives the following

$$\begin{pmatrix} mC + (1\pm 1)V - \varepsilon & \mp\dfrac{\mu}{\kappa}m + \dfrac{\kappa}{\mu}CV - \dfrac{d}{dr} \\ \mp\dfrac{\mu}{\kappa}m + \dfrac{\kappa}{\mu}CV + \dfrac{d}{dr} & -mC + (1\mp 1)V - \varepsilon \end{pmatrix} \begin{pmatrix} \phi^+ \\ \phi^- \end{pmatrix} = 0, \qquad (4)$$

where $C = \sqrt{1-(\mu/\kappa)^2}$. Since $|\kappa| = 1, 2, ...$, then for real solutions we require that $|\mu| \leq 1$. Now, Eq. (4) results in the following relation between the two spinor components

$$\phi^{\mp} = \frac{1}{mC \pm \varepsilon}\left(\frac{d}{dr} \pm \frac{\kappa}{\mu}CV - \frac{\mu}{\kappa}m\right)\phi^{\pm}, \tag{5}$$

and the following second order differential equation

$$\left[-\frac{d^2}{dr^2} + \frac{\kappa}{\mu}C\left(\frac{\kappa}{\mu}CV^2 \mp V'\right) + 2\varepsilon V + m^2 - \varepsilon^2\right]\phi^{\pm} = 0, \tag{6}$$

which is Schrödinger-like. Equations (5) and (6) with the top/bottom sign are valid for $\varepsilon \neq \mp mC$, respectively. Since $C > 0$, then Eq. (5) and (6) with the top/bottom sign are valid ONLY for positive/negative energy. We should emphasize that Eq. (6) will NOT give the two spinor components belonging to the same energy space. One has to choose one sign in Eq. (6) to obtain ONLY one of the two components then substitute that into Eq. (5) with the corresponding sign to obtain the other component. The positive and negative energy subspaces are disconnected. This is a general feature of the solution space of the Dirac equation, which is unfortunately overlooked more often than not. Now, we can obtain an approximate analytic solution of the problem for all $\kappa$ if the Schrödinger-like equation (6) is exactly solvable in the presence of $V$, $V^2$, and $V'$. The second and third terms in Eq. (6) resemble the two potential partners $V_\pm = \mathcal{W}^2 \pm \mathcal{W}'$ in supersymmetric quantum mechanics [9], where $\mathcal{W} = \frac{\kappa}{\mu}CV$. However, supersymmetry is spoiled by the contribution of the $\varepsilon V$ term. As examples, we consider the Hulthén potential, $V_0/(e^{\lambda r} - 1)$, and the 1/r singular part of the Eckart potential, $V_0/\tanh(\lambda r)$, where $\lambda > 0$. For small $\lambda$ both potentials behave near the origin like $(V_0/\lambda)r^{-1}$. Therefore, in both cases $\mu = V_0/\lambda$ with $|V_0| \leq \lambda$. We will solve the problem for positive energy. That is, we take the top signs in Eqs. (5) and (6). The negative energy solutions could simply be obtained by applying the following map on the positive energy solutions (both the energy spectrum and the spinor wavefunctions)

$$\phi^{\pm} \to \phi^{\mp}, \; \varepsilon \to -\varepsilon, \; \kappa \to -\kappa, \; V_0 \to -V_0. \tag{7}$$

We start by solving the Dirac-Hulthén problem. If we define the variable $x = e^{-\lambda r} \in [0, +1]$, then the positive energy Schrödinger-like equation (6) in the new variable $x$ reads as follows

$$\left[-x^2\frac{d^2}{dx^2} - x\frac{d}{dx} - \frac{\gamma x}{(1-x)^2}(\gamma x + 1) + \frac{2\varepsilon V_0}{\lambda^2}\frac{x}{1-x} + \frac{m^2 - \varepsilon^2}{\lambda^2}\right]\phi^+ = 0, \tag{8}$$

where $\gamma = \kappa C = \kappa\sqrt{1 - (V_0/\lambda\kappa)^2}$. We take the ansatz $\phi^+(r) = x^\alpha(1-x)^\beta F(x)$, where $\alpha$ and $\beta$ are real positive parameters. This turns Eq. (8) into the differential equation for the hypergeometric series with $F(x) = {}_2F_1(a,b;c;x)$ provided that

$$\alpha = \lambda^{-1}\sqrt{m^2 - \varepsilon^2}, \text{ and } \beta = \begin{cases} \gamma + 1 & , \kappa > 0 \\ -\gamma & , \kappa < 0 \end{cases}. \tag{10}$$

Therefore, real solutions are possible only for $|\varepsilon| < m$ (i.e., bound states). The parameters of the hypergeometric function become

$$c = 2\alpha + 1, \; a = \alpha + \beta - \sqrt{\alpha^2 + \gamma^2 - 2\varepsilon V_0/\lambda^2}, \; b = \alpha + \beta + \sqrt{\alpha^2 + \gamma^2 - 2\varepsilon V_0/\lambda^2}. \tag{11}$$

Reality of these parameters require that the energy spectrum be bound by the condition

$$(\varepsilon + V_0)^2 \leq m^2 + \lambda^2\kappa^2. \tag{12}$$

Moreover, bound state solution requires that the hypergeometric series terminate. This means that either $a$ or $b$ should be a negative integer. It is clear that $b$ can never be negative. Thus, we require that $a = -n$, where $n = 0,1,2,..$, giving $b = n + 2(\alpha + \beta)$. This also results in the condition that $(n + \alpha + \beta)^2 = \alpha^2 + \gamma^2 - 2\varepsilon V_0/\lambda^2$ whose solution gives the following positive energy spectrum formula

$$\varepsilon_n = \frac{\lambda}{2}\left[\left(\frac{V_0}{\lambda}\right)^2 + (n+|\gamma|)^2\right]^{-1}\left\{-\left(\frac{V_0}{\lambda}\right)n(n+2|\gamma|) + (n+|\gamma|)\right.$$
$$\left.\times\sqrt{\left(\frac{2m}{\lambda}\right)^2\left[n(n+2|\gamma|) + \kappa^2\right] - n^2(n+2|\gamma|)^2}\right\}, \qquad (13)$$

where $n = 0,1,2,...,n_{max}$ and $n_{max}$ is found from (12) with $\varepsilon = \varepsilon_{n_{max}}$ or, equivalently, from the reality of the square root in (13). It is calculated as the largest integer, which is less than or equal to the following number

$$N = |\gamma|\left\{-1 + \sqrt{1 + 2\left(\frac{m}{\lambda\gamma}\right)^2\left[1 + \sqrt{1 - (\lambda\kappa/m)^2}\right]}\right\}. \qquad (14)$$

We should note that the energy eigenvalue associated with the spinor wavefunction $\psi_n = \begin{pmatrix} \psi_n^+ \\ \psi_n^- \end{pmatrix}$ is $\varepsilon_n$ for $\kappa < 0$ and $\varepsilon_{n+1}$ for $\kappa > 0$. The lowest positive energy is obtained from Eq. (13) for $n = 0$ giving $\varepsilon_0 = mC = m\gamma/\kappa$. Now, since no results exist in the literature on this problem to compare our findings with, we look at special cases. Taking $\kappa = -1$ in the spectrum formula (13) gives the S-wave solution of the Dirac-Hulthén problem [10]. We should note that there are many studies on the relativistic S-wave Hulthén problem, but almost all impose either the "spin" or "pseudo-spin" symmetry [4]. Those have energy spectra that differ from the proper Dirac-Hulthén S-wave problem, like the one given by Eq. (3.13) in [10]. On the other hand, the nonrelativistic limit is obtained by taking $\varepsilon \approx m + E$, where $|E| \ll m$ and $|V_0| \ll m$. A better and more natural way to obtain this limit is by putting back the $\hbar$ and $c$ in (13) and letting $c \to \infty$. Keeping the lowest order terms in $c^{-1}$ gives $\varepsilon \approx mc^2 + E$ and results in the following nonrelativistic energy spectrum for the Hulthén potential with non-zero angular momentum [11]

$$E_n^\ell = -\frac{\lambda^2}{8}\left[\frac{2(V_0/\lambda^2) - \ell^2}{n + \ell + 1} + (n + \ell + 1)\right]^2, \qquad (15)$$

where we let $n \to n+1$ in (13) since $\kappa = \ell > 0$. Moreover, for ease of comparison with results in the literature, we wrote (15) in the atomic units $\hbar = m = 1$. The energy spectrum formula given in [12] by Eq. (14) agrees with the above but only if we replace $\ell^2$ by $\ell(\ell+1)$. On the other hand, Eq. (19) for the energy spectrum in [13] agrees with the above if we take their adjustable numerical parameter $\omega = 0$ and replace $\ell(\ell+1)$ by $\ell^2$. However, the $\ell^2$ term is completely missing from the spectrum formula (28) of Ref. [14]. Now, the S-wave ($\ell = 0$) restriction of (15) reproduces the well-known nonrelativistic exact result [1].

Using the definition of the Jacobi polynomials $P_n^{(\mu,\nu)}(z)$ in terms of the hypergeometric series [15], we can write the upper spinor component as follows

$$\phi_n^+(r) = \begin{cases} A_n^{\gamma+1}e^{-\lambda\alpha_{n+1}r}(1-e^{-\lambda r})^{\gamma+1}P_n^{(2\alpha_{n+1},2\gamma+1)}(1-2e^{-\lambda r}) & ,\kappa > 0 \\ A_n^{-\gamma}e^{-\lambda\alpha_n r}(1-e^{-\lambda r})^{-\gamma}P_n^{(2\alpha_n,-2\gamma-1)}(1-2e^{-\lambda r}) & ,\kappa < 0 \end{cases} \qquad (16)$$

where $A_n^{\gamma+1}$ is a normalization constant and $\alpha_n = \lambda^{-1}\sqrt{m^2 - \varepsilon_n^2}$. The lower component is obtained from this by using Eq. (5) with the top sign and utilizing the differential formula and recursion relations of the Jacobi polynomials [10,15]. Finally, the negative energy solutions (energy spectrum and wavefunction) are obtained from these positive energy solutions by applying on them the simple map (7).

Now, we turn briefly to the second example, the Dirac-Eckart problem, and give results without too many details. We write Eq. (6) in terms of the variable $x = \coth(\lambda r) \in [+1,\infty]$ and take the upper component as $\phi^+(r) = (x+1)^{-\alpha}(x-1)^{\beta} F(x)$. Boundary conditions dictate that $\alpha > \beta > 0$. The result is, again, a differential equation for $F(x) = {}_2F_1(a,b;c;\frac{1-x}{2})$ provided that

$$\alpha = \tfrac{1}{2}\sqrt{\kappa^2 + \lambda^{-2}\left[m^2 - (\varepsilon+V_0)^2\right]},\ \beta = \tfrac{1}{2}\sqrt{\kappa^2 + \lambda^{-2}\left[m^2 - (\varepsilon-V_0)^2\right]}. \tag{17}$$

For bound states, the hypergeometric series terminates with its parameters as follows
$$a = -n,\ b = 2(\beta-\alpha)+n+1,\ \text{and}\ c = 2\beta+1. \tag{18}$$

The energy spectrum formula is obtained from the condition $n+\beta-\alpha = \begin{cases} -\gamma-1, & \kappa>0 \\ \gamma, & \kappa<0 \end{cases}$ as

$$\varepsilon_n = \lambda\left[1+(V_0/\lambda)^2\left(n+|\gamma|\right)^{-2}\right]^{-1/2}\left[(m/\lambda)^2 - n(n+2|\gamma|)\right]^{1/2}, \tag{19}$$

where $n = 0,1,2,...,n_{max}$ and $n_{max}$ is found from the reality condition of the energy spectrum as the largest integer that is less than or equal to

$$N = |\gamma|\left[-1 + \sqrt{1+(m/\lambda\gamma)^2}\right]. \tag{20}$$

Here again, the energy eigenvalue associated with the spinor wavefunction $\psi_n(r)$ is $\varepsilon_n$ for $\kappa < 0$ and $\varepsilon_{n+1}$ for $\kappa > 0$. The upper spinor component reads as follows

$$\phi_n^+(r) = \begin{cases} A_n^{\gamma+1}(x+1)^{-\alpha_{n+1}}(x-1)^{\beta_{n+1}} {}_2F_1(-n, 2\beta_{n+1}-2\alpha_{n+1}+n+1; 2\beta_{n+1}+1; \frac{1-x}{2}), & \kappa > 0 \\ A_n^{-\gamma}(x+1)^{-\alpha_n}(x-1)^{\beta_n} {}_2F_1(-n, 2\beta_n-2\alpha_n+n+1; 2\beta_n+1; \frac{1-x}{2}), & \kappa < 0 \end{cases} \tag{21}$$

where $\alpha_n$ and $\beta_n$ are obtained from (17) with $\varepsilon \to \varepsilon_n$. The condition above Eq. (19) shows that $\alpha_n - \beta_n > n$, which guarantees that $\lim_{r \to 0}\phi_n^+(r) = 0$. The lower component is obtained from this by using Eq. (5) with the top sign and utilizing the differential formula of the hypergeometric function [15]. Now, all results reported in the literature about the relativistic energy spectrum of the Eckart potential are not concerned with the pure vector coupling mode presented in this work. As stated above, almost all investigations are done with spin or pseudo-spin symmetry imposed on the problem where the spinor couples to a combined scalar and vector Eckart potential. Nonetheless, we can compare the non-relativistic limit of our result in (19) with studies that obtain approximate analytic evaluation of the energy spectrum of the Eckart potential in the Schrödinger equation with non-zero angular momentum [16]. Taking the nonrelativistic limit ($c \to \infty$) of (19) and using the atomic units, $\hbar = m = 1$, we obtain

$$E_n^\ell = -\frac{\lambda^2}{2}\left[\left(\frac{V_0/\lambda^2}{n+\ell+1}\right)^2 + (n+\ell+1)^2 - \ell^2\right]. \tag{22}$$

With $\ell = 0$, this formula reproduces the exact and well-know S-wave energy spectrum for the nonrelativistic problem.

In conclusion, we have managed to obtain, for the first time, an approximate analytic solution (energy spectrum and spinor wavefunction) for the Dirac equation with coupling to pure 1/r singular *vector* potentials for all angular momenta. The major contribution of our approach is in the approximation that is made for $r^{-1}$, which is less singular than the traditional $r^{-2}$. The latter is widely used in the conventional approach but with the limitation of accuracy to the lowest part of the energy spectrum. The less singular approximation used here makes it possible to extend the validity of the results to higher excitation levels in the spectrum. As an illustration of the utility of the approach, we treated the Dirac-Hulthén and the Dirac-Eckart problems.

**Acknowledgements:** The support of this work by the Saudi Center for Theoretical Physics (SCTP) is highly appreciated. We acknowledge partial support by King Fahd University of Petroleum & Minerals (KFUPM).